\documentclass[12pt,dvips]{article}
\usepackage{graphicx}
\usepackage{amsmath}
\usepackage[psamsfonts]{amssymb}
\usepackage{amsthm}
\usepackage{euscript}

\usepackage{latexsym}

\renewcommand{\theequation}{\arabic{section}.\arabic{equation}}
\renewcommand{\(}{\begin{equation}}
\renewcommand{\)}{end{equation} \vspace{-.05in}\linebreak}

\newcounter{saveeqn}
\newcounter{savealpheqn}

\newcommand{\alpheqn}{\setcounter{saveeqn}{\value{equation}}%
 \stepcounter{saveeqn}\setcounter{equation}{0}%
 \renewcommand{\theequation}{\mbox{\arabic{section}.\arabic{saveeqn}\alph{equation}}}
 \renewcommand{\)}{\end{equation}}}
\def\part#1{\frac{\partial}{\partial{#1}}}%

\def\group#1{\refstepcounter{equation}\setcounter{saveeqn}{\value{equation}}%
\label{#1}\setcounter{equation}{0}%
\renewcommand{\theequation}{\mbox{\arabic{section}.\arabic{saveeqn}\alph{equation}}}%
\renewcommand{\)}{\end{equation}}}

\newcommand{\reseteqn}{\setcounter{equation}{\value{saveeqn}}%
 \renewcommand{\theequation}{\arabic{section}.\arabic{equation}}%
 \renewcommand{\)}{\end{equation}}}

\newcommand{\aalpheqn}{\setcounter{saveeqn}{\value{equation}}%
 \stepcounter{saveeqn}\setcounter{equation}{0}%
 \renewcommand{\theequation}{\mbox{\Alph{subsection}.\arabic{saveeqn}\alph{equation}}}
  \renewcommand{\)}{\end{equation}}}
\newcommand{\che}{{\tilde{H}}^e}
\newcommand{\cho}{{\tilde{H}}^o}
\newcommand{\he}{{\tilde{H}}_e}
\newcommand{\ho}{{\tilde{H}}_o}
\newcommand{\areseteqn}{\setcounter{equation}{\value{saveeqn}}%
 \renewcommand{\theequation}{\Alph{subsection}.\arabic{equation}}%
 \renewcommand{\)}{\end{equation}}}

\renewcommand{\thefootnote}{\alph{footnote}}
\renewcommand{\(}{\begin{equation}}
\renewcommand{\)}{\end{equation}}
\newcommand{\ba}{\begin{eqnarray}}
\newcommand{\ea}{\end{eqnarray}}

\newcommand{\bp}{\mathop{\vtop{\ialign{##\crcr
  $\hfil\displaystyle{}\hfil$\crcr\noalign{\kern-13pt\nointerlineskip}
  \BIG{(}\hskip0pt\crcr\noalign{\kern3pt}}}}}
\newcommand{\cbp}{\mathop{\vtop{\ialign{##\crcr
  $\hfil\displaystyle{}\hfil$\crcr\noalign{\kern-13pt\nointerlineskip}
  \BIG{)}\hskip0pt\crcr\noalign{\kern3pt}}}}}
\newcommand{\pa}{\mathop{\vtop{\ialign{##\crcr
  $\hfil\displaystyle{\oplus}\hfil$\crcr\noalign{\kern+1pt\nointerlineskip}
  \hspace{.08in}$^{\alpha=0}$\hskip6pt\crcr\noalign{\kern3pt}}}}}
\renewcommand{\sp}{,\hspace{.3in}}
\newcommand{\p}{^\prime}

\newcommand{\R}{\ensuremath{\mathbb R}}

\newcommand{\Z}{\ensuremath{\mathbb Z}}

\newcommand{\beq}{\begin{equation}}
\newcommand{\eeq}{\end{equation}}

\newcommand{\rp}{\mathbf{RP}}

\numberwithin{equation}{section}

\catcode`\@=11
\def\vereq#1#2{\lower3pt\vbox{\baselineskip1.5pt \lineskip1.5pt
\ialign{$\m@th#1\hfill##\hfil$\crcr#2\crcr\sim\crcr}}}
\catcode`\@=12

\makeatletter
\newcommand\tabcaption{\def\@captype{table}\caption}
\makeatother
\renewcommand{\(}{\begin{equation}}
\renewcommand{\)}{\end{equation}}

\begin{document}

\begin{titlepage}
\begin{flushright}
hep-th/0211172
\end{flushright}

\vspace{2em}
\def\thefootnote{\fnsymbol{footnote}}

\begin{center}
{\Large  IIB Soliton Spectra with All Fluxes Activated}
\end{center}
\vspace{1em}

\begin{center}
Jarah Evslin\footnote{E-Mail: jarah@df.unipi.it} 
\end{center}

\begin{center}
\vspace{1em}
{\em INFN Sezione di Pisa\\
     Universita di Pisa\\
     Via Buonarroti, 2, Ed. C,\\
     56127 Pisa, Italy}\\
\end{center}

\vspace{3em}
\begin{abstract}
\noindent

\end{abstract}
Building upon an earlier proposal for the classification of fluxes, a sequence is proposed which generalizes the AHSS by computing type IIB string theory's group of conserved RR and also NS charges, which is conjectured to be a K-theory of dual pairs. As a test of this proposal, the formalism of Maldacena, Moore and Seiberg (hep-th/0108100) is applied to classify D-branes, NS5-branes, F-strings and their dielectric counterparts in IIB compactified on a 3-sphere with both NS and RR background fluxes.   The soliton spectra on the 3-sphere are then compared with the output of the sequence, as is the baryon spectrum in Witten's non-$spin^c$ example, $AdS^5\times\rp^5$.  The group of conserved charges is seen to change during Brown-Teitelboim-like phase transitions which change the effective cosmological constant.  
\vfill
November 17, 2002

\end{titlepage}
\setcounter{footnote}{0} 
\renewcommand{\thefootnote}{\arabic{footnote}}

\pagebreak
\renewcommand{\thepage}{\arabic{page}}
\pagebreak 

\section{Introduction}
\group{dummy}\reseteqn

There is no known S-duality covariant classification of strings and branes nor of fluxes in type IIB string theory.  The supergravity equations of motion, Bianchi identities and gauge equivalences indicate that if we consider fluxes to be differential forms then we should classify them by de Rham cohomology.  Applying an exact sequence of Moore and Witten \cite{MW}, this would suggest that charges should similiarly be classified by de Rham cohomology.   While this may yield the correct answer in the supergravity theory, the quantum theory is complicated by the Dirac quantization condition.  This suggests that solitons instead be classified by integral homology as in Ref.~\cite{baryonz}.  However in Ref.~\cite{baryonz} it is seen that the set of allowed charges is only a subset of integral homology.  This is understood as the effect of a global worldsheet anomaly in Ref.~\cite{FW}.  In fact, as first observed in Ref.~\cite{DMW} and later formalized in Ref.~\cite{MMS}, this anomaly implies that the group of conserved charges consists of equivalence classes of subsets of integral homology.

These equivalence classes of subsets are produced by the Atiyah-Hirzebruch spectral sequence (AHSS) which, up to an extension problem, computes the K-theory of spacetime.  This is reassuring as it has been conjectured \cite{MM} that D-branes in IIB are classified by the K-group $K^0$.  More generally, as conjectured in Ref.~\cite{Ktheory,BandM}, a non-trivial background $H$-field leads to a classification by the twisted K-group $K_H^0$. 

A classification of NS charges by homology and RR charges by K-homology does not respect the S-duality covariance which is conjectured to exist in IIB.  In particular such a classification cannot be correct because the formalism of Ref.~\cite{MMS} does respect S-duality and implies that NS charges are also classified by equivalence classes of subsets of integral cohomology \cite{Uday}.  In addition, the S-dual of the Myers dielectric effect \cite{Myers} indicates that there is also some nontrivial extension problem to be solved for the NS fields before they can be classified, an extension problem which is likely to involve both the NS and RR sectors.  It is unknown what classification will result\footnote{One may hope that the self-duality of the RR fluxes would be manifest in such a formulation.  This may imply that K-theory of infinite-dimensional vector bundles be replaced by something with less degrees of freedom, such as a classification of dimensionally-reduced $E_8$ bundles.} from the solution of this extension problem, nor whether it will ultimately be S-duality covariant \cite{AP}.  

The purpose of this note is to calculate, at weak string coupling\footnote{We do not attempt to understand the extent to which the presence of NS5-branes violates this assumption.  However the S-duality arguments below are consistent with this assumption as they only rely upon the SL(2,$\Z$)-invariant structure of the WZW terms in various actions and not on the coupling of the dilaton.}, the spectra of D-branes, NS5-branes and F-strings in IIB string theory compactified on the product of a 7-manifold and a 3-sphere.  At different points in the paper different retrictions will be imposed on the kinds of 7-manifolds allowed.  We compare the results against those obtained by the S-duality covariant extension of the AHSS in Ref.~\cite{Uday}.  This extension calculates a generalization of K-cohomology and classifies fieldstrengths, whereas in the present note we wish to classify solitons.  Therefore we modify the extension so that it calculates a generalization of K-homology, where the solitons are conjectured to live.  

In Section~\ref{exsec} we calculate the soliton spectrum of IIB on the 3-sphere.  In Sec.~\ref{revsec} we review the AHSS and its physical interpretation by Maldacena, Moore and Seiberg (MMS).  We also review the S-duality covariant extension which is not a spectral sequence but is a set of maps between cohomology groups reminescent of the AHSS.  The physical processes involved will already be familiar from the computation on the 3-sphere.  Next, in Sec.~\ref{conjsec} we will create the homology version of the extended AHSS for use with $spin^c$ spacetime manifolds.  This is well suited for computation, although it has explicit dependence on background fields.  The consistency of such an explicit dependence and its physical interpretation will be discussed, including the fact that the group of conserved charges changes when a physical process changes the background fields\footnote{The old charge is still conserved but no longer counts branes, until such a time as the background fields are changed back.}.  

Section~\ref{specsec} extends the previous ideas to include D9-branes and in particular explains, as indicated in \cite{DMW}, that the Sen conjecture is a special case of the MMS processes considered in this paper.  This is used to generate some outlandish speculations about the nature of the new K-theory.  In Sec.~\ref{compsec} we compute the soliton spectrum of IIB on the 3-sphere using the homology sequence and compare the result to the direct computation of Section~\ref{revsec}.  This requires the use of a variation of homology which agrees with the usual homology (singular with $\Z$ coefficients) on a compact manifold but in general encodes, for example, information about the brane content on the horizon of $AdS$.  We conclude with a conjectured extension to the non-$spin^c$ case in Sec.~\ref{nonspincsec}, which is tested against the baryon spectrum on $AdS^5\times\rp^5$.

\section{Branes on the Three-Sphere} \label{exsec}
Consider type IIB string theory on any open, contractible 7-manifold crossed with a 3-sphere.  By ``crossed'' we mean homeomorphic to the Cartesian product, but worldsheet conformal invariance is likely to demand that the 10-dimensional metric not be a total product.  The 3-sphere is taken to be large, although in the case of D-brane wrappings with only NS flux it was seen in Ref.~\cite{FS} that many of the solitons discussed below continue to exist when this approximation is relaxed, that is, on the SU(2) WZW model at finite level.  $j$ units of $G_3$ RR 3-form flux and $k$ units of $H$ NS 3-form flux are supported on the sphere.  $j$ and $k$ are not necessarily responsible for keeping the sphere large, this may be accomplished by boundary conditions.  We will restrict attention to branes with the following property, which we believe contains a brane of every possible charge.  Let $M_x$ be the intersection of a brane with the $S^3$ at a point $x$ in the 7-manifold\footnote{Where $x$ is defined by some homeomorphism of the 10-manifold to a Cartesian product space.}.  Then all $M_x$'s that are not the empty set must be homeomorphic to each other.  In other words, each brane will be, topologically, a product of a submanifold of the 7-manifold and a submanifold of the 3-sphere.  We will also restrict our attention to the case in which both $j$ and $k$ are nonvanishing except when explicitly stated otherwise.

Clearly there may be any number of D-instantons, D-strings, D3-branes and F-strings which are completely unwrapped on the (have a 0-dimensional intersection with each) three-sphere.  The unwrapped 5-branes are domain walls.  In particular $j$ or $k$ changes by 1 when crossing a D$5$ or NS$5$-brane respectively.  Whether such a shift is acceptable depends on the kind of boundary conditions imposed. General configurations of D$7$-branes are not S-duality covariant \cite{DMW} but fortunately do not fit in the non-compact directions here\footnote{However a collection of coincident D5-branes, due to the H-field background, will generically blow up into a dielectric D7-brane.  This carries no net D7-brane charge and appears consistent with S-duality, although such an assertion may lead to the existence of some $(p,q)$ 7-brane \cite{Ftheory} that carries NS5-brane charge in the presence of a $G_3$ flux.}.  Notice, for example, that an NS7-brane (defined as a source for the dilaton) is not consistent with our small coupling approximation because, in the presence of a nonvanishing axion field, the dilaton acquires a nontrivial monodromy around loops which link the NS7-brane.

\begin{figure}[ht] 
  \centering \includegraphics[width=5.5in]{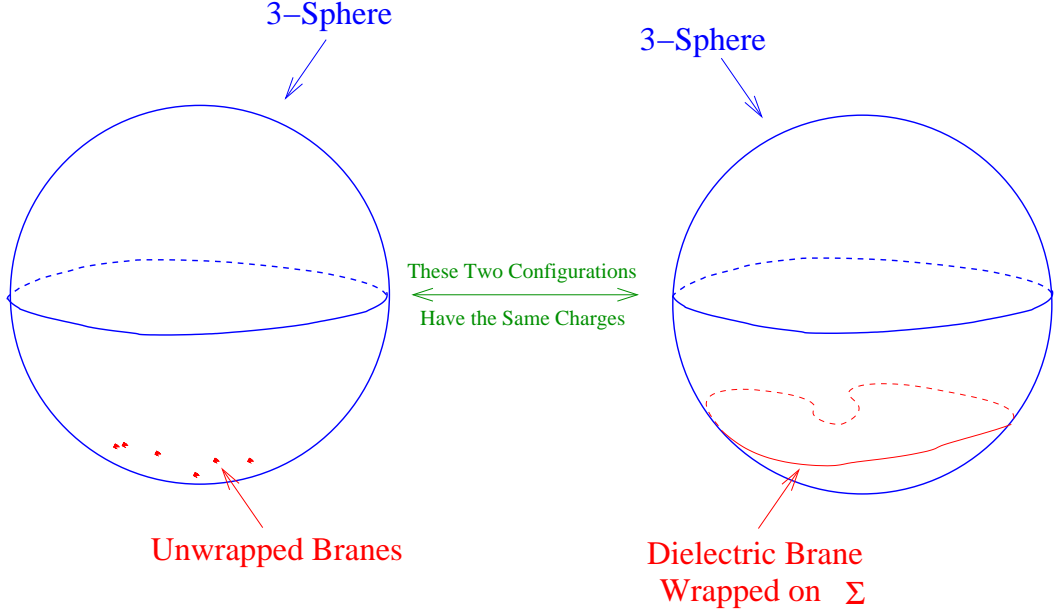}
\caption{$k$ branes which each intersect an $S^3$ at a point carry the same charge as one brane which intersects an $S^3$ at a manifold $\Sigma$ if the integral of the first Chern class of its gauge bundle over $\Sigma$ is equal to $k$.  A paralell to the Kondo model indicates \cite{FS} that not only do these two configurations carry the same charges, but also that there are dynamical processes which interpolate between them.} \label{myerfig}
\end{figure}

Less trivial is the classification of objects that wrap some $p$-dimensional submanifold of the 3-sphere.  Only the 0 and 3 dimensional homology groups of $S^3$ are nonvanishing.  This implies that any brane wrapping a submanifold of dimension $p<3$ can be deformed to a point (crossed with some submanifold of the 7-fold) and so carries conserved charges that may be carried by a pointlike brane.  However in the last paragraph we have already classified charges whose intersection with each $S^3$ is a discrete set of points, thus $p<3$ cannot lead to any new charges.  The charge of an object wrapping such a submanifold is obtained by an application of S-duality to the observation of Refs.~\cite{Taylor,Polchinski} that if a D$p$-brane with gauge fieldstrength $F$ is wrapped on a contractible manifold $\Sigma$ then it carries D$p$ and D$(p-2)$ brane charges
\begin{equation} \label{wati}
Q_{Dp}=\int_\Sigma c_0(F)\sp
Q_{D(p-2)}=\int_\Sigma c_1(F)
\end{equation}
where $c_i(F)$ is the $i$th Chern class of a gauge bundle with fieldstrength $F$.  Notice that D$p$-brane charge is only carried for $p=0$, the D$p$-branes that do not wrap the 3-sphere at all.  

If $p=1$ then both integrals vanish.  However the brane may still carry lower-dimensional brane charges.  In IIB this adds no additional complication as a slight perturbation of such a brane can be made such that it intersects each $S^3$ in a discrete set of points and then this case is reduced to $p=0$. The case $p=2$ is illustrated in Figure~\ref{myerfig} and is the subject of the next subsection.

\subsection{Branes Wrapping 2d Submanifolds}

Subcases of $p=2$ have been studied extensively in the literature and consist of dielectric branes.  Although these configurations only carry the charges of lower-dimensional branes obtained by collapsing $\Sigma$ to a point, when the background fluxes are nonvanishing the potential energy of the dielectric brane is generally minimized by a $\Sigma$ with positive area.  For example in the case of the supersymmetric SU(2) WZW at level $k$ ($k$ units of $H$ flux) it was calculated in Ref.~\cite{BDS} that the minimum energy configuration is a 2-sphere at latitude $(Q/k-1/2)\pi$.  Notice (\ref{wati}) that although the strength of the background $H$-flux determines the size and shape of the dielectric brane, the charges of the dielectric branes are independent of this flux and depend only on the topology of the worldvolume gauge bundle.  Thus to compute the charges of dielectric branes we need only consider the worldvolume gauge couplings, which are S-duality covariant.  S-duality may then be used to find the entire spectrum of branes wrapping 2-dimensional submanifolds $\Sigma\subset S^3$ and their conserved charges.  We will now use Eq.~(\ref{wati}) and S-duality to find the spectrum of dielectric branes one dimension at a time.

A D-instanton does not have enough dimensions to wrap $\Sigma$.  A ($p,q$)-string may wrap $\Sigma$, but then must be instantonic and so does not carry a conserved charge.  A D3-brane wrapping $\Sigma$ carries dielectric (p,q)-string charge \cite{Tamar} equal to
\begin{equation}
q=Q_{F1}=\int_{\Sigma}c_1(*F)\sp
p=Q_{D1}=\int_{\Sigma}c_1(F).
\end{equation}
If this D3 is instantonic then its decay to a fundamental string has been described in Ref.~\cite{John}.  D5 and NS5-brane wrappings yield the same D3-charge, because magnetic monopoles in both worldvolume theories are D3-brane boundaries:
\begin{equation} \label{d3}
Q_{D3}=\int_{\Sigma}c_1(F).
\end{equation}
More generally Eq.~(\ref{d3}) holds for the U(1) fieldstrength of the worldvolume gauge theory of any $(p,q)$ 5-brane with $p$ and $q$ relatively prime.  Finally a wrapped D7-brane may carry D5-brane charge
\begin{equation} 
Q_{D5}=\int_{\Sigma}c_1(F).
\end{equation}
It would be tempting to extend this formula to describe NS5-brane charge in some 7-brane, but this would require a suspicious application of S-duality.  Furthermore if the 7-brane is an NS7-brane it violates our weak coupling assumptions, whereas a D7 would be counterintuitive because it would be much lighter than the original NS5 if $g_s$ is taken to zero before the radius of the $S^3$ is taken to infinity.

The charges just defined are the Page charges of Ref.~\cite{Marolf}.  As discussed there  $\int F$ is not invariant under large gauge transformations, and so the physical charges are the gauge orbits of the charges calculated above.  Thus to compute the group of charges we need to calculate the effect of a large gauge transformation on $\int F$ and quotient by it.  To do this, we relate $F$ to a quantity which is gauge invariant, $F+B$, and then use Stoke's theorem to compute the integral of $B$ by integrating $H$ over some three-manifold bounded by $\Sigma$.  The choice of 3-manifold is precisely the choice of gauge.  Choosing a different 3-manifold changes the integral of $B$ and, since $B+F$ is well defined, the integral of $F$ must change as well.  Thus we may calculate the gauge transformation of $\int F$: 
\begin{eqnarray} \label{don}
2\pi Q_{\textup{\small{invar}}}&=&\int_\Sigma (F+B)=\int_\Sigma F + \int_{M^3} H=2\pi Q_{\textup{\small{Page}}} + \int_{M^3} H\nonumber\\&=&2\pi Q_{\textup{\small{Page}}}\p - \int_{S^3 - M^3} H=2\pi Q_{\textup{\small{Page}}}\p - 2\pi k +\int_{M^3} H
\end{eqnarray}
where $M^3$ is a submanifold of $S^3$ whose boundary is $\Sigma$.  Thus a large gauge transformation shifts the charge by a multiple of $k$, the NS 3-form flux on the 3-sphere
\begin{equation}
Q_{\textup{\small{Page}}}\longrightarrow Q_{\textup{\small{Page}}}\p=Q_{\textup{\small{Page}}}+k.
\end{equation}
Quotienting by this gauge transformation we learn that the unwrapped D-brane charges are classified by at most $\Z_k$ for each kind of quantum number that describes the D-branes in the noncompact space\footnote{For example such a D-brane might end on a boundary or horizon of the noncompact space and get quantum numbers from there.}.  

More ambiguities are caused by the $j$ units of $G_3$ flux supported on the $S^3$.  Applying S-duality to the background, which induces a Montonen-Olive duality on a D3-brane worldvolume gauge theory, one finds that although $\int *F$ is not gauge invariant the quantity $\int (*F+C_2)$ is gauge invariant.  The above argument then yields an ambiguity in F-string charge, it may shift by $j$ units.  Thus the F-strings with each type of boundary condition in the contractible space are classified by $\Z_j$.  The gauge invariant quantity on the worldvolume theory of an NS5-brane is $\int (F+C_2)$, and so the D3-brane charge, in addition to the above ambiguity of shifts by $k$ units, has another ambiguity of shifts by $j$ units.  Thus D3-brane charge is classified by $\Z_{gcd(j,k)}$.  Similarly D5-branes are classified by at most $\Z_k$, but we are hesitant to discuss contributions from other 7-branes such as that from D7-branes in the presence of a $G_3$ flux.

The brane may also wrap a compact contractible submanifold $M$ in the contractible 7-manifold.  However by integrating $ch_{dim(M)/2}$ one can find a scenario with the same charges but a brane extended in dim($M$) less dimensions that is trivial in the contractible space.

If we relaxed the contractibility condition on our 7-manifold then, in addition to wrapping the compact manifold $\Sigma$, the brane in question may also wrap a nontrivial cycle in the 7-manifold.  This leads to new quantum numbers, but as will be seen in Sec.~\ref{compsec} we may treat these quantum numbers just as the ones above that described boundary conditions in a contractible space.

\begin{figure}[ht] 
  \centering \includegraphics[width=5.5in]{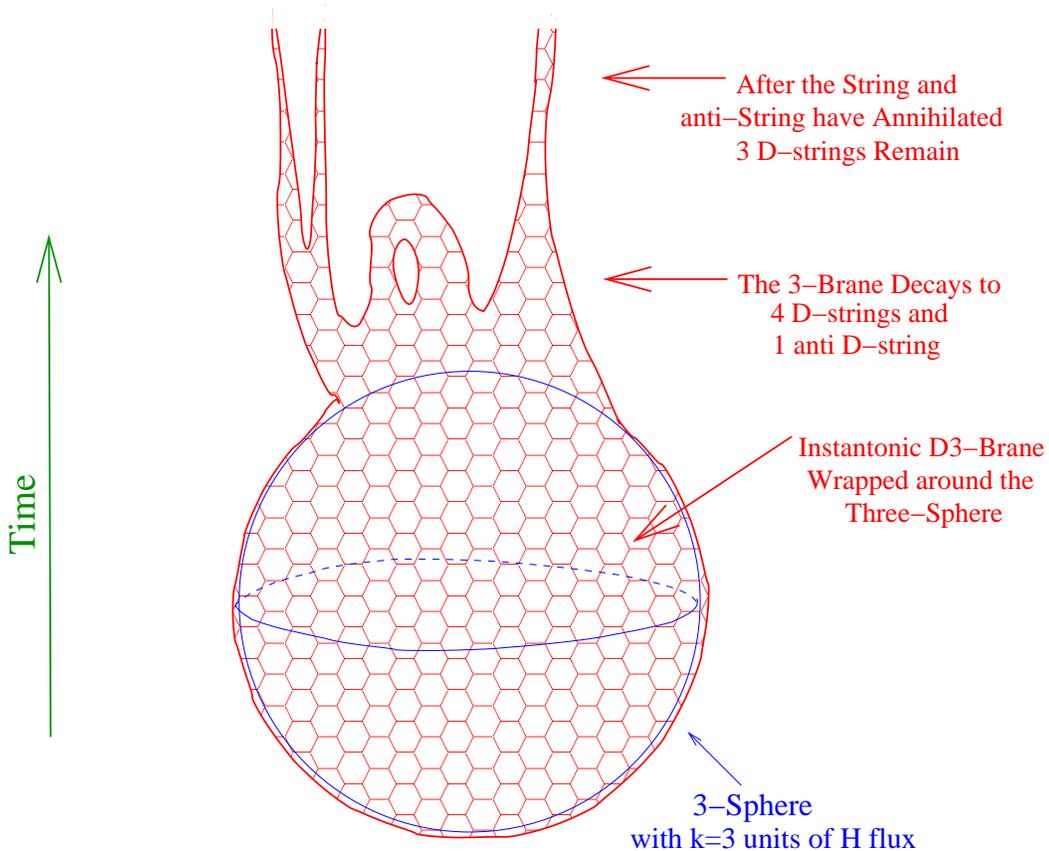} 
\caption{A D3-brane wraps a 3-sphere with $k$ units of $H$-flux and no $G_3$ flux.  It also extends in a spacelike direction which is not shown.  The vertical direction of the diagram is timelike and at each moment in time a crosssection $\Sigma$ of the $S^3$ is chosen on which the D3 is wrapped.  As $\Sigma$ shrinks to a point if we use the gauge choice which keeps $B$ finite then $\int_\Sigma F=6\pi$ and so the D3 cannot decay entirely.  Instead it decays to 4 D-strings and 1 anti D-string.  A D-string and an anti D-string annihilate, leaving 3 D-strings.  Thus D-string number is only conserved modulo $3$ while F-string number is conserved completely.} \label{decayfig}
\end{figure}

\subsection{Branes Wrapping the Whole 3-Sphere}

There are two kinds of branes wrapping 3-submanifolds of the 3-sphere.  The first wraps a contractible 3-submanifold.  To study such a brane we may introduce a Morse function on the 3-submanifold and perturb the brane slightly so that the preimages of different points of the Morse function lie at different points on the 7-manifold.  Then for each point on the 7-manifold the embedded surface is a 2-dimensional level surface $\Sigma$ and we are reduced to the case studied above.  An explicit construction of such a brane as one of the parafermionic branes of $\cite{MMS}$ has appeared in Refs.~\cite{Sarkis,Sarkis2}.

Finally we will classify objects wrapped around the entire 3-sphere.  Such an object must be at least 3-dimensional.  For example a D3-brane may wrap the 3-sphere, leaving one direction unaccounted for.  If this direction is spacelike then the D3-brane is an example of one of the instantonic D-branes of Ref.~\cite{MMS}.  In particular its decay violates F-string charge by $j$ units and D-string charge by $k$ units.  Alternately one may consider a Morse function on the $S^3$ to be the time direction\footnote{Such a parametrization violates the ``product of two submanifolds'' condition above and so was not considered in the previous subsection.}, in which case (ignoring the non-wrapping direction) the D3 begins as a point, expands through some family of $\Sigma$'s and then turns into a point and disappears.  To keep $(B,C_2)$ finite at the beginning and the end one must do a gauge transformation during the lifetime of this brane which shifts $(c_1(F),c_1(*F))$ by $(k,j)$ units.  This changes the $(p,q)$-string charge by $(k,j)$ units and so at the end of the D3's lifetime it cannot disappear altogether, but instead leaves a $(k,j)$ string, just as indicated by the analaysis of Ref.~\cite{MMS}.  This means that the charge group of $(p,q)$-strings is $\Z_k\times\Z_j$, as found above in the equivalent analysis of branes wrapping 2-cycles.  An example with the RR flux turned off may be seen in Figure~\ref{decayfig}.

\begin{figure}[ht] 
  \centering \includegraphics[width=5.5in]{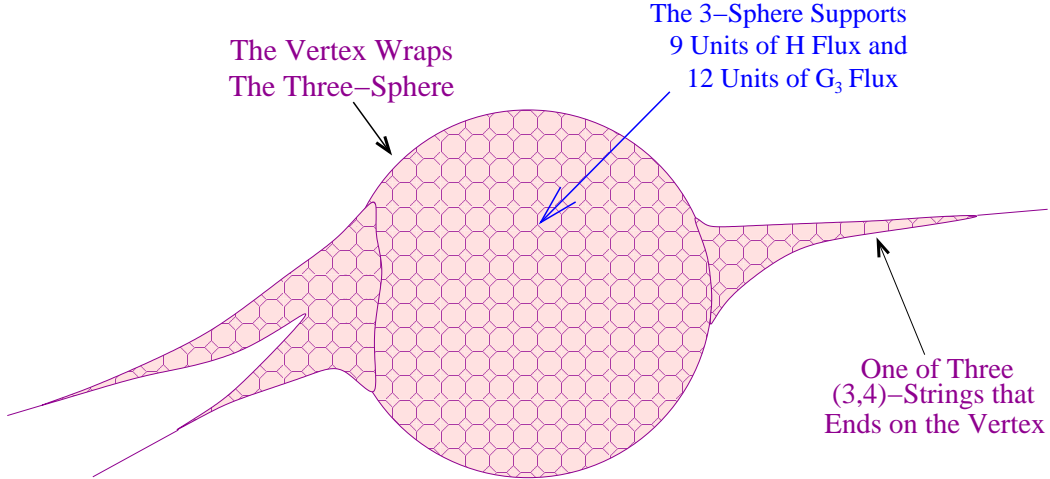} 
\caption{One instant in the life of a D3-brane (vertex) wrapping a 3-sphere with $k=9$ units of $H$-flux and $j=12$ of $G_3$ flux.  It also extends in a timelike direction which is not shown.  The induced gauge theory fluxes cannot exist on a compact space and so the D3 must have a noncompact worldvolume.  In this picture it has noncompact tendrils with string charges.  In a ground state in flat space there will be $3=gcd(9,12)$ tendrils each of which carries $(3,4)$-string charge.  The width of a tendril is inversely proportional to its distance from the vertex \cite{HanWit}, and so these strings are always slightly dielectric.}\label{baryonfig}
\end{figure}

If instead the extra direction is timelike then the D3 will be a lower-dimensional version of the baryon vertex of Ref.~\cite{baryonz}, although an interpretation as a vertex would require the use of probe 5-branes surrounding the noncompact directions whose worldvolume gauge theory provides charges for the quark fields.  In particular, a D3-brane wrapping the $S^3$ and extending in a timelike direction must provide one of the endpoints of $j$ F-strings and $k$ D-strings.  Dynamically the energy of $j$ F-strings and $k$ D-strings is minimized if they form $gcd(j,k)$ separate boundstates of $(k/gcd(j,k),j/gcd(j,k))$-strings \cite{bound} each of which has a worldvolume theory which is S-dual to that of a F-string.  Thus the configuration leads to a vertex which couples $gcd(j,k)$ dyons in a probe theory.  Alternately, one can consider this object which wraps the $S^3$ and has $j$ free F-string ends and $k$ free D-string ends to be one of the solitons of the theory.  We will call this a vertex because the first example led to a baryonic vertex in the dual gauge theory.  An example may be seen in Figure~\ref{baryonfig}. 

One may also wrap a D5-brane on the 3-cycle.  If the D5 is instantonic then the $H$ flux implies that the decay product includes $k$ D3-branes.  The role of the $G_3$ flux is more mysterious, as the D5 worldvolume action contains the coupling
\begin{equation}
S\supset C_2\wedge F\wedge F
\end{equation}
and so naively the decay products include $j$ 3+1 dimensional objects with a worldvolume action term $F\wedge F$, or instanton number, with a theta-angle aparently equal to $2\pi$.  Whether such exotic dielectric strings exist or not depends on the dynamics of the theory, here we are only interested in charges, and so instead will search for a more conventional source for the same charge.

\begin{figure}[ht] 
  \centering \includegraphics[width=5.5in]{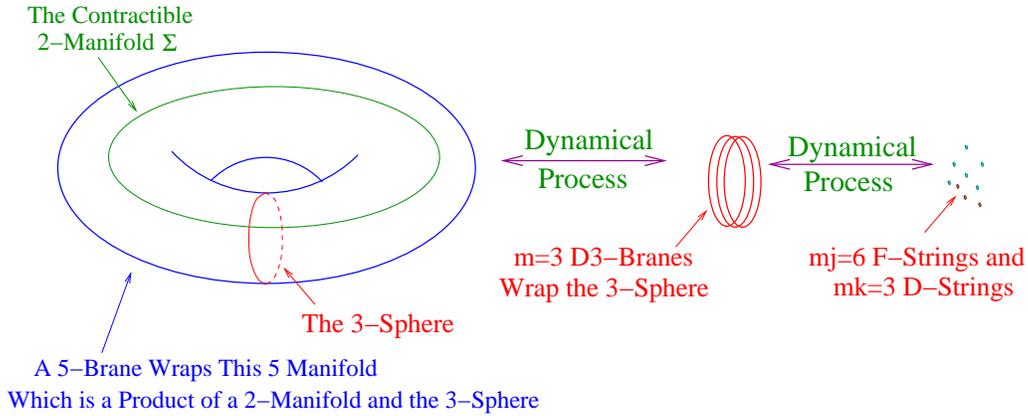} 
\caption{A D5-brane wraps the $S^3$, a contractible 2-cycle $\Sigma$ and some other irrelevant direction.  The integral of the first Chern class of the field strength over $\Sigma$ is $m=3$.  The 3-sphere supports $j=2$ units of $G_3$ flux and $k=1$ of $H$ flux.  In this case the D5-brane carries the same charges as $m=3$ D3-branes which wrap $S^3$ but not $\Sigma$.  It also has the same charges as $jm=6$ D-strings and $km=3$ F-strings, which in a supersymmetric background one might expect to form $m\times gcd(j,k)=3$ independent boundstates each of which is a $(k/gcd(j,k),j/gcd(j,k))=(1,2)$-string.}\label{doublefig}
\end{figure}

A D5-brane or NS5-brane wrapped around the 3-sphere occupies some 3-dimensional manifold in the other directions.  Choose a Morse function on this manifold.  The preimage of each point is a 2-dimensional surface $\Sigma$.  As above, one may integrate the 2-form gauge fieldstrength $F$ over $\Sigma$.  The result (divided by $2\pi$) is the D3 charge on $\Sigma$.  The 5-brane may then be allowed to decay in two steps.  The first gives a D3, whose worldvolume can be Poincare dual of $F$ in the 5-brane worldvolume.  This D3-brane still wraps $S^3$, but now the problem has been reduced to the problem solved above, although in this reduction one must remember that whenever $\Sigma$ is noncompact the 3-brane must be given all of the charges that the 5-brane receives from boundary conditions.  Thus the wrapped instantonic 5-brane may decay to wrapped instantonic D3's which then decay as above to a $(k,j)$-string that does not wrap the $S^3$ but instead is extended in time and also is extended along some curve homologous to the PD of $F$ in the projection of the 5-brane worldvolume onto the contractible dimensions.  Such double-staged decays have been seen previously \cite{MMS} in IIA string theory on the SU(3) group manifold.  An example may be seen in Figure~\ref{doublefig}.

Similarly a wrapped 7-brane may be replaced with a wrapped 5-brane which is dual to its fieldstrength, so long as the boundary charges can remain unchanged.  If this is impossible then, as with the case of the 5-brane, the replacement is not valid and the decay product is not understood.  For example if there is a noncontractible $T^2$ in the 7-dimensional space then a D5 may wrap $T^2\times S^3$ and also another direction.  This yields a ``not understood'' decay.  T-dualizing both cycles of the torus leads to an understood decay of a D3 wrapped on the 3-sphere.  The decay product is a $(k,j)$ string, and so our mysterious decay product is related by two T-dualities to a $(k,j)$ string.  When $k\neq 0$ this is just a $k$ D3-branes with $j$ units of worldvolume magnetic flux.  If $k=0$ then this is a collection of strings T-dualized twice, and so the decay product is a collection of strings whose endpoints on the D5 are Poincare dual to $F\wedge G_3$.

If the 5-brane is not instantonic then it yields yet another kind of vertex, but this time the vertex extends in two more dimensions.  Again, in the case of the 5-brane, D-strings coming out of the old vertex are replaced by D3-branes here, and again the substitute for the F-string is the same mysterious-looking extended object which consists of ordinary strings when the D5 can decay to a D3.  Obstructions to such a decay live on the boundary of space (and in the topology of the 7-fold when we allow it to be nontrivial) where the obstructed D5 ends, and the resolution to what becomes of the string is related to the nature of this boundary.  Similarly D7-branes yield a kind of vertex that extends in $4+1$ dimensions.

\subsection{Summary}

We have seen that, except for the vertices, all configurations of branes and strings in this $M^7\times S^3$ carry the same charges as some collection of branes and strings on $M^7$ alone.  The role of the $S^3$ is to violate the conservation of some of these charges.  This is because the dynamical process of an object being shoved around the $S^3$ changes the charges by a discrete amount $n$, whose values depend on the fluxes of the $S^3$.  The result is that all of the charges (except for possibly the vertices) take values in finite cyclic groups $\Z_n$. 

For each kind of quantum number associated with $M^7$ the charges of strings and branes on $M^7\times S^3$ are summarized in Table 1.  Dielectric versions of each object are also listed with a formula for computing their lower-dimensional charge.  Only the dielectric branes which will be relevant to the study of $d_3$ that follows are included.

\vspace{.3in}
\hspace{1in}
\begin{tabular}{c|c|c|c}
Charge&Charge&Charge&Formula\\
Type&Group&Carrier&for Charge\\ \hline
F1&$\Z_j$&D3&$\int *F$\\
D1&$\Z_k$&D3&$\int F$\\
D3&$\Z_{gcd(j,k)}$&D5,NS5&$\int F$\\
D5&$\Z_k$&D7&$\int F$\\
Vertex&$\Z$&N/A&N/A\\
\multicolumn{4}{c}{}\\
\multicolumn{4}{c}{Table 1: Charges of Dielectric Branes on $S^3$}
\end{tabular}

\vspace{.3in}

\section{Review: The AHSS and S-duality Covariant AHSS} \label{revsec}
\subsection{The AHSS and the MMS Prescription}

The Atiyah-Hirzebruch spectral sequence (AHSS) is an algorithm for computing, up to an extension problem, the twisted K-cohomology of a given manifold.  We will be interested in only the first step of this sequence, which on some backgrounds (such as $AdS^5\times S^5$) will yield a group of brane charges which includes some anomalous branes and fails to identify the charges of some configurations that are related by dynamical processes \cite{MMS}.  

First define a differential operator $d_3$ on the cohomology ring $H^*$ of spacetime by
\begin{equation}
d_3:H^p\longrightarrow H^{p+3}:x\mapsto (Sq^3+H)x
\end{equation}
where $Sq^3$ is a Steenrod square and $H$ is the background H fieldstrength.  The first approximations to the K-cohomology groups are then
\alpheqn
\begin{equation}
K_H^0\sim \oplus_j \frac{ker(d_3:H^{2j}\longrightarrow H^{2j+3})}{im(d_3:H^{2j-3}\longrightarrow H^{2j})}.
\end{equation}
\begin{equation}
K_H^1\sim \oplus_j \frac{ker(d_3:H^{2j+1}\longrightarrow H^{2j+4})}{im(d_3:H^{2j-2}\longrightarrow H^{2j+1})}.
\end{equation}
\reseteqn
It has been conjectured \cite{MM,Ktheory} that D-branes in IIB string theory with a background H flux are classified by $K_H^0$ of the 10-dimensional spacetime manifold.  If the spacetime manifold splits into $\R\times M^9$ where $\R$ represents the time direction, then one may speak of conserved quantities and it has been conjectured \cite{MMS} that the correct classification group is roughly $K_H^1(M^9)$. 

As explained in Ref.~\cite{Ktheory}, if a RR flux is not annihilated by $d_3$ then the D-brane that created that flux necessarily has a worldvolume on which $W_3+H$ does not vanish, where $W_3$ is the third Stieffel-Whitney class of its normal bundle.  Such a D$p$-brane suffers from the Freed-Witten anomaly \cite{FW} and is inconsistent unless D$(p-2)$-branes end on it such that the intersection is Poincare dual to $W_3+H$ in $p$-brane worldvolume.  Such a process was seen in Figure~\ref{baryonfig}.  Therefore if we try to classify states that do not involve branes ending on branes, we find that the only allowed charges are those in the kernel of $d_3$.  Later we will see that the sequence approach discards exactly the states in our $S^3$ example that involve branes ending on branes, the vertices.

We have seen that the charges are restricted to be valued in $ker(d_3)$.  However as discussed in Ref.~\cite{MMS}, these charges are violated by dynamical processes such as that in Fig.~\ref{decayfig} where an instantonic-brane is inflicted with the Freed-Witten anomaly and so lower-dimensional branes are emitted before the instantonic brane self-destructs.  Recall that to cure the anomaly the endpoints of these lower-dimensional branes need to be dual to $W_3+H$, but these endpoints are a slice of constant time and so determine the charge carried.  Thus the charge is roughly that of $W_3+H$ times that of the other directions wrapped by the instantonic brane.  More precisely this ``times'' is a cup product with the Poincare dual of the instanton, and $W_3$ needs to be pushed forward onto the spacetime manifold, which yields $Sq^3$ of the Poincare dual\footnote{This is only true up to corrections which appear in the higher order differentials.}.  Thus the resulting brane charge is $Sq^3+H$ times the charge of the instantonic brane.  It is then easy to believe that this process has violated charge by an element of the image of $d_3$, and so charges are preserved at best modulo $im(d_3)$.  A first approximation to the charges of D-branes in an H background is then $ker(d_3)/im(d_3)$.

Through a converging sequence of successive approximations, beginning with (\ref{step1}), the AHSS yields the set of conserved charges.  However to find the group structure on this set one must next solve an extension problem.  In the S-duality invariant case below the form of this extension problem is not known.  However in examples such as IIB on $S^3$ above, the correct group structure is apparent from the physics\footnote{While this construction does appear to classify D-branes with nothing ending on them, it may be noted that each of these instantonic processes changes the field content.  This change must be ignored to arrive at the K-theory classification.  But one may then be suspicious of the K-theory classification of fluxes, as it was derived by dualizing a classification of charges in which flux shifts were explicitly dropped.  In fact, if they were not dropped and if fluxes corresponding to vertices were allowed one might believe that RR fluxes were classified by ordinary cohomology.  We will defer such cynicism to a sequel.}.

\subsection{The S-duality Covariant AHSS}
In Ref.~\cite{Uday} this prescription was partially extended to one that yields both NS and RR fieldstrengths in IIB.  To do this, one first defines the odd part of a generalized cohomology of the space, whose equivalence classes are gauge inequivalent NS and RR fieldstrengths.  This consists of the tensor product of every odd cohomology group with the SL$(2,\Z)$ module corresponding to the S-duality representation inhabited by fields of that dimension.  Explicitly 
\begin{equation} \label{step1}
\cho=H^1\oplus H^3\oplus H^3\oplus H^5\oplus H^7\oplus H^7.
\end{equation}
The even part of the generalized cohomology is the tensor product of each cohomology group of dimension $2k$ with $SL(2,\Z)$ module corresponding to the S-duality representation inhabited by Poincare duals of $(9-2k)$-branes
\begin{equation} 
\che=H^2\oplus H^4\oplus H^6\oplus H^8\oplus H^8\oplus H^{10}.
\end{equation}
A better understanding of dielectric NS5-branes may lead to a second $H^4$ term.



Now we can define the generalized ``differentials'' (which are non-linear and so not actually differentials) to be maps
\begin{equation}
d_3: \cho \longrightarrow \che.
\end{equation}
In the Moore-Witten \cite{MW} interpretation of the relation between charges and fluxes an element of $\cho$ characterizes the field configuration on a homologically trivial domain wall such as a sphere at infinity.  Such a configuration topologically characterizes the brane content in the region bounded by the domain wall.  In particular the field configuration determines the Freed-Witten anomaly on the branes that created it.  This anomaly is cancelled by other branes which necessarily intersect the domain wall.  $d_3$ takes a field configuration on the domain wall to the Poincare dual of this intersection of the anomaly-cancelling branes and the wall.  This is why $\che$ consists of Poincare duals of branes.  

For example, consider the spacetime $AdS^5\times S^5$.  The domain wall may be the product of $S^5$ with the horizon of $AdS^5$ and the flux configuration may dictate that there is a vertex inside.  In this case the quarks are F-strings which end on the horizon and the image of a differential\footnote{In this case the relevant differential is not $d_3$ but rather $d_5$.} is the Poincare dual of the intersection of these quarks with the horizon.  

To be consistent with this interpretation the formulas above must not use the cohomology of the 10-dimensional spacetime, but rather the cohomology of the 9-dimensional boundary.  This does not affect of the dimensions of the fluxes and branes because the fluxes created by objects in the interior run along the domain wall (That is, their projection to the domain wall is a form of the same dimension.) while the $(9-2k)$-branes restricted to a wall are one dimension lower, but the Poincare duality is performed in 9 dimensions instead of 10 and so the resulting cohomology class is still of dimension $2k$.

We will first restrict our attention to $spin^c$ spacetime manifolds so that the Steenrod square terms in the generalized AHSS are trivial.  In this case the differential acts by \cite{Uday}
\group{cahss}
\begin{equation}
d_3(G_1)=H\cup G_1
\end{equation}
\begin{equation}
d_3(G_3,H)=H\cup G_3
\end{equation}
\begin{equation}
d_3(G_5)=(H\cup G_5,G_3\cup G_5)
\end{equation}
\begin{equation}
d_3(G_7=*G_3,H_7=*H)=H\cup G_7.
\end{equation}
\reseteqn
Here $G_p$, the RR p-form flux, is any element of $H^p$ while $H$ and $*H$ are elements of $H^3$ and $H^7$ respectively. 

The fascinating thing about the generalized spectral sequence is that it is not a spectral sequence at all, but rather it contains products of the fieldstrengths and so is highly nonlinear.  This is as it must be, because in the classical limit the sequence enforces the supergravity equations of motion for the fieldstrengths, which are themselves highly nonlinear.  

The physical configurations on 10-dimensional spacetime (or its 9-dimensional boundary) consist of the kernel of this map.  However, in this note we are not interested in the classification of field configurations on the 10-dimensional spacetime, but rather in the set of field configurations on each 9-dimensional timeslice modulo the physical processes which relate them.  These are the conserved quantities.

No perscription has been given for how to group the elements of the kernel into such equivalence classes except in the case in which there are no background RR fluxes and one only tries to classify D-branes.  However the Maldacena, Moore, Seiberg (MMS) procedure of $\cite{MMS}$ suggests that two field configurations in $\cho$ have the same charges (are in the same equivalence class) if they differ by an element of the image of some unknown map $d_3\p$ from even to odd dimensional cohomology.  If\footnote{This may separate the cohomology into two Hodge-dual pieces, one of which relates to physical charges and so is quantized.  It was argued in Ref.~\cite{TopEff} that quantized fluxes are Hodge-dual to non-quantized fluxes.}  spacetime factors into $\R\times M^9$ then the relevant cohomology classes are the cohomology of an $8$-dimensional domain wall in $M^9$.  The fluxes created by instantonic branes in $M^9$ will extend in the time direction and so on the $8$-dimensional domain wall will restrict to fluxes of dimension one lower, that is, of even dimension.  $d_3$ then acts on these even dimensional fluxes.  The arguments of \cite{Uday} may be used to create the unknown map, although it will depend on background odd-dimensional fluxes as a result of the nonlinearity of this system.  We will perform such a construction explicitly for the homology version of this sequence shortly.

\section{The Conjectured Sequence for Homology} \label{conjsec}

\subsection{The Generalized Homology Groups}

Our goal is to compare the classication of charges on the 3-sphere with the spectra predicted by the S-duality covariant generalization of the AHSS.  There are several obstructions to making such a comparison.  Among these the most fundamental is that the generalized spectral sequence does not yield a classification of charges (generalizing K-homology) but instead yields a classification of fields (generalizing K-cohomology).  In Ref.~\cite{MM}, Moore and Witten find an exact sequence which relates these classifications in the case in which there are no background RR fluxes and also no NS5-branes and fundamental strings.  Intuitively one treats the $H$ field as a background while taking the exterior derivative of the RR fields.  After this everything must be Poincare dualized to get results about a homology theory rather than a cohomology theory.

In the present case, as we are trying to classify NS5-branes and F-strings, the $H$-field is not a constant background field but instead is on equal footing with $G_3$ as a source.  Thus we will need to apply the procedure of Moore and Witten in a more symmetric fashion.  A natural guess at a symmetric generalization of Moore and Witten's prescription is that instead of taking the exterior derivative of only the RR fluxes, one takes the exterior derivative (for torsion terms this must be interpretted as a coboundary operation) of the entire AHSS differential, and then Poincare dualizes.  This will be made explicit momentarily.

First one defines generalized even and odd homology by tensoring with the SL(2,$\Z$) modules inhabited by the charges
\begin{equation}  \label{hom}
\he=H_2\oplus H_2\oplus H_4\oplus H_6\oplus H_6\oplus H_8\sp
\ho=H_1\oplus H_1\oplus H_3\oplus H_5\oplus H_5\oplus H_7. 
\end{equation}
These groups are easier to interpret than the cohomology groups above.  Employing the MMS viewpoint, the even homology groups $H_p$ describe $(p+0)$-dimensional instantonic branes, such as instantonic D$(p-1)$-branes.  The odd homology groups $H_p$ are the charges of objects of dimension $p+1$, such as static D$p$-branes.  To enforce such a viewpoint we will require that our spacetime splits into $\R\times M^9$ where $\R$ is identified with the time direction.  By a conserved charge we then mean a nontrivial wrapping in the 9 spatial directions, and so the charges of interest are homology classes of $M^9$.  This implies that we may restrict our attention to the classification of static branes without losing generality.

In general charges valued in $H^*$, $H_*$, $K^*$ or $K_*$ of the 10-dimensional spacetime manifold are those classes which are invariant under some class of deformations.  On the other hand, charges classified by the topology of a 9-dimensional Cauchy surface are said to be conserved if they are independent of the Cauchy surface chosen, that is, if they are not changed by time-dependent physical processes.  The formulae for these two types of invariants are very similar because often the time-dependent physical processes are the 10d deformations with time as the deformation parameter and one dimension omitted.  However the 9$d$ charges are easier to manipulate because the deformation parameter is time and so may be treated as a physical direction, for example branes may be extended along it.  For this reason the construction of the map $d_3\p$, as we will soon see, is trivial in the second picture.    

\subsection{The ``Differentials''}

The differential $d_3$ takes a static brane to its intersection with the branes that cure its Freed-Witten anomaly.  Thus it is a map
\begin{equation} \label{d3h}
d_3:\ho \longrightarrow \he
\end{equation}
where we recall that the charge of a static brane is the homology class in $\ho$ of a time-slice.  The classification of sources is then a set of equivalence classes of the kernel of $d_3$, which consists of the set of allowed configurations.  Again the MMS procedure suggests that we construct equivalence classes by quotienting by the image of some map
\begin{equation}
d_3\p:\he \longrightarrow \ho.
\end{equation}
The physical charge group of the branes is then
\begin{equation}
\textup{Charges}=\frac{ker(d_3)}{im(d_3\p)}.
\end{equation}
So what are the maps $d_3$ and $d_3\p$?

As mentioned above, we will define $d_3$ so that, roughly, the right hand side is the exterior derivative of the right hand side of (\ref{cahss}).  This is only true roughly because some terms are more naturally included in the higher differential $d_5$ in this approach instead of $d_3$, such as those describing an NS5-brane with $G_5$ flux in its worldvolume and therefore D-string charge. 

We will denote the two NS charges and the RR charges in $H_p$ by $F_1$, $NS_5$ and $D_{p-1}$ respectively.  The conjectured action of $d_3$ on homology is then
\group {homseq}
\begin{equation}
d_3(D_7)=D_7 \cap PD(H)
\end{equation}
\begin{equation}
d_3(D_5,NS_5)=D_5 \cap PD(H) + NS_5\cap PD(G_3)
\end{equation}
\begin{equation}
d_3(D_3)=(D_3 \cap PD(H),D_3 \cap PD(G_3))
\end{equation}
\begin{equation}
d_3(D_1,F_1)=0.
\end{equation}
\reseteqn
$PD(x)$ is the homology class Poincare dual to the cohomology class $x$.  An extension to the non-$spin^c$ case is conjectured in Eq.~(\ref{guess}).

Although much easier to interpret, this differential is more mysterious than the cohomology differential of Eq.~(\ref{cahss}) because it involves $H$ and $G_3$ as separate inputs, whereas they may be sourced by the very objects being classified.  This is a consequence of the non-linearity of the supergravity equations of motion from which it was derived.  Due to this bizarre dependence, the consistency of this approach depends highly on the benevolent asymptotics of the $M^p\times\R$ spacetime and will be analyzed below.

A conserved charge corresponds to a $(2k+1)$-brane wrapping a $(2k+1)$-cycle in the spatial manifold $M^9$ and also extended in the time direction.  Thus the conserved charges are equivalence classes of elements of $ker(d_3)\subset\ho$.  The equivalence classes are determined by quotienting the set of allowing configurations, $ker(d_3)$, by the image of 
\begin{equation}
d_3\p:\he\longrightarrow\ho
\end{equation}
which consists of fixed-time slices of configurations that may be created from nothing via instantonic processes.  Now we will use the splitting of space-time into space and time to define $d_3\p$.

$d_3\p$, like $d_3$, simply computes the Poincare dual of the relevant 3-form on a fixed-time slice of the worldvolume, the only difference is that $d_3\p$ acts on the worldvolume of a $p$-brane with all $p+1$ dimensions lying along the slice.  Thus $d_3\p$ is defined by precisely the same Eq.~(\ref{homseq}) as $d_3$, the only difference is that the homology classes $D_p$, $F_1$ and $NS_5$ are now defined to be elements of $H_{p+1}$, $H_2$ and $H_6$ respectively. 

\subsection{The Consistency of Background Fieldstrengths}
Finally we turn to the interpretation of the explicit dependence on fieldstrengths of Eq.~(\ref{homseq}).  Intuitively, in keeping with the picture of Moore and Witten, we would like these to be boundary values of the fieldstrengths.  However we have allowed $M^9$ to be an arbitrary $spin^c$ manifold, and so there is not necessarily a sphere at spatial infinity.  There is a distant hypersurface which is comprised of the far past and far future.  With this motivation we define the ``background'' fields $G_3$ and $H$ to be those on an $M^9$ at some arbitrary time and study the consistency of such a choice.  

If there are any 5-branes in this timeslice then the cohomology class of these background fields is not well defined, because a 5-brane is a domain wall across which a 3-form fieldstrength jumps.  For simplicity let us first consider the case in which there are no 5-branes on this slice.  Now there is no problem defining the $G_3$ and $H$ at this moment in time and then plugging them into the above differentials.  However the consistency of this approach needs to be investigated because 5-brane bubbles with no net charge may spontaneously form and inside of them the 3-form fluxes will be different and so the instantonic brane decay processes will yield different products, yielding a different classification of charges inside the bubble.  Thus we must check that, while the charge classification may have changed locally inside the bubble, the charges integrated over all of $M^9$ remain conserved in the presence of such bubbles. 

\begin{figure}[ht] 
  \centering \includegraphics[width=5.5in]{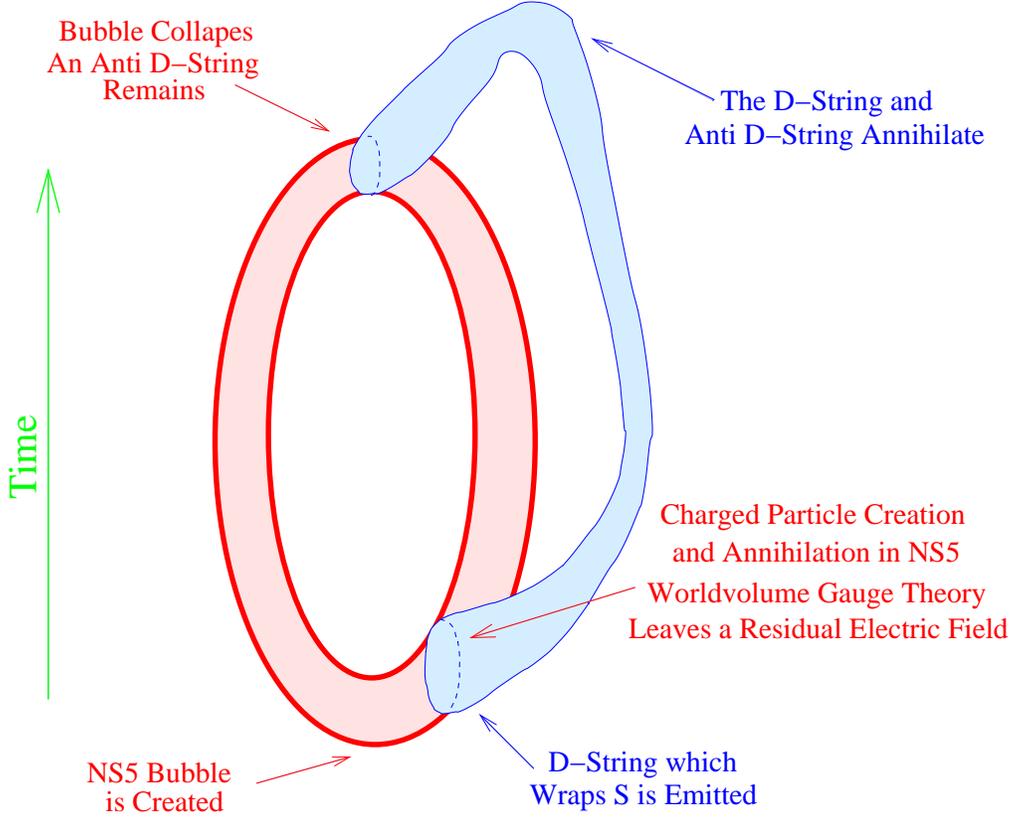}
\caption{The First Process: An instantonic NS5-brane bubble wrapping $S$ is spontaneously created and emits a D-string which also wraps $S$.  From the perspective of the 6-dimensional worldvolume gauge theory an electrically charged particle/anti-particle pair have been created, wound around $S$ and then annihilated, leaving an electric field.  The electric field would prevent the NS5 bubble from collapsing and so the inverse process, emission of an anti D-string wrapping $S$, occurs before the NS5 bubble collapses.  In this example the D-string and anti D-string later annihilate.  This process conserves D-string charge.}\label{scen1}
\end{figure}

More concretely, we need to worry about the following 2 processes and their S-duals.  Consider a background $H$ flux and a 3-cycle $Z^3$ such that
\begin{equation}
\int_{Z^3}\frac{H}{2\pi}=k.
\end{equation}
$H$ may be a torsion class in which case we recall that such integrals are defined by the homology-cohomology pairing.  Also there may be other nontrivial cycles around, as well as nontrivial $G_3$ flux.  An instantonic 3-brane may wrap $Z^3$ and another 1-cycle $S\in H_1(M^9)$ and then decay to $k$ D-strings which wrap $S$ and are also extended in time.  Thus D-strings wrapping $S$ are classified by some quotient of $\Z_k$.  If this quotient is the trivial group then there will be no conservation laws and so no inconsistencies, and so to search for potential inconsistencies we will consider the case in which this quotient is nontrivial.  Now we will create an instantonic NS5 bubble and try to use it to violate the conservation of D1 charge.  The NS5-brane must wrap $S$ or this will be impossible.  The two scenarios that could potentially violate D-string conservation are as follows.

In the first process, illustrated in Figure~\ref{scen1}, the NS5-brane may emit a D-string wrapped around $S$.  This certainly has changed the number of D-strings wrapped around $S$ by 1 unit, which is nontrivial in the relevant charge group.  The number of D-strings wrapping $S$ is then not a conserved quantity.  However the NS5-brane enjoys a 6-dimensional worldvolume U(1) gauge theory.  In this gauge theory the endpoint of the D-string is an electric charge source with a worldline which is $S$.  Thus this emission of a D-string corresponds, in the worldvolume gauge theory, to the creation of an electrically charged particle and antiparticle from nothing, their relative voyage around a noncontractible loop, and their subsequent annihilation.  If you end your life with a nonzero winding number, you will be remembered (and furthermore the universe will be unable to end until you are resurrected (and unwind)).

After the annihilation no electric charge remains in the gauge theory, however the noncontractibility of the loop $S$ implies that there still is a residual $*F$ electric flux.  Therefore in order to form a conserved quantity for D-string charge wrapped around $S$ one must consider an NS5-brane with $*F$ flux to also carry this charge.  Notice that if the NS5-brane bubble recollapses then this $*F$ charge forces it to emit an anti D-string wrapped around $S$, cancelling the charge of the wrapped D-string that it created earlier.  Thus the combined charge is conserved by this process in which an NS5 bubble is created, expands, contracts and disappears.  This argument respects S-duality as the worldvolume gauge couplings are S-duality covariant, that is, it proceeds identically for D5-branes and F-strings with a background $G_3$ flux.  More precisely the induced fields are the results of the following couplings on the 3 and 5-brane worldvolumes:
\begin{equation} \label{coups}
S_{D3}\supset *F\wedge B+F\wedge C_2\sp
S_{D5}\supset *F\wedge B\sp
S_{NS5}\supset *F\wedge C_2 .
\end{equation}
The S-duality argument only relies on the S-duality invariance of Eq.(\ref{coups}).

A similar argument also applies to D3-branes  emitted from 5-branes.  In this case the D3-brane's end is a magnetic source in the 6-dimensional gauge theory.  The source decays but if the D3 wraps a nontrivial 3-cycle then so does the magnetic source and so a gauge fieldstrength $F$ remains on the worldvolume after the source has disappeared.  This magnetic field carries D3 charge and if the 5-brane bubble decays this $D3$ charge will remain, in the form of an anti D3-brane wrapping the same cycle as the D3-brane emitted earlier.  Thus the total D3 charge is conserved.

\begin{figure}[ht] 
  \centering \includegraphics[width=5.5in]{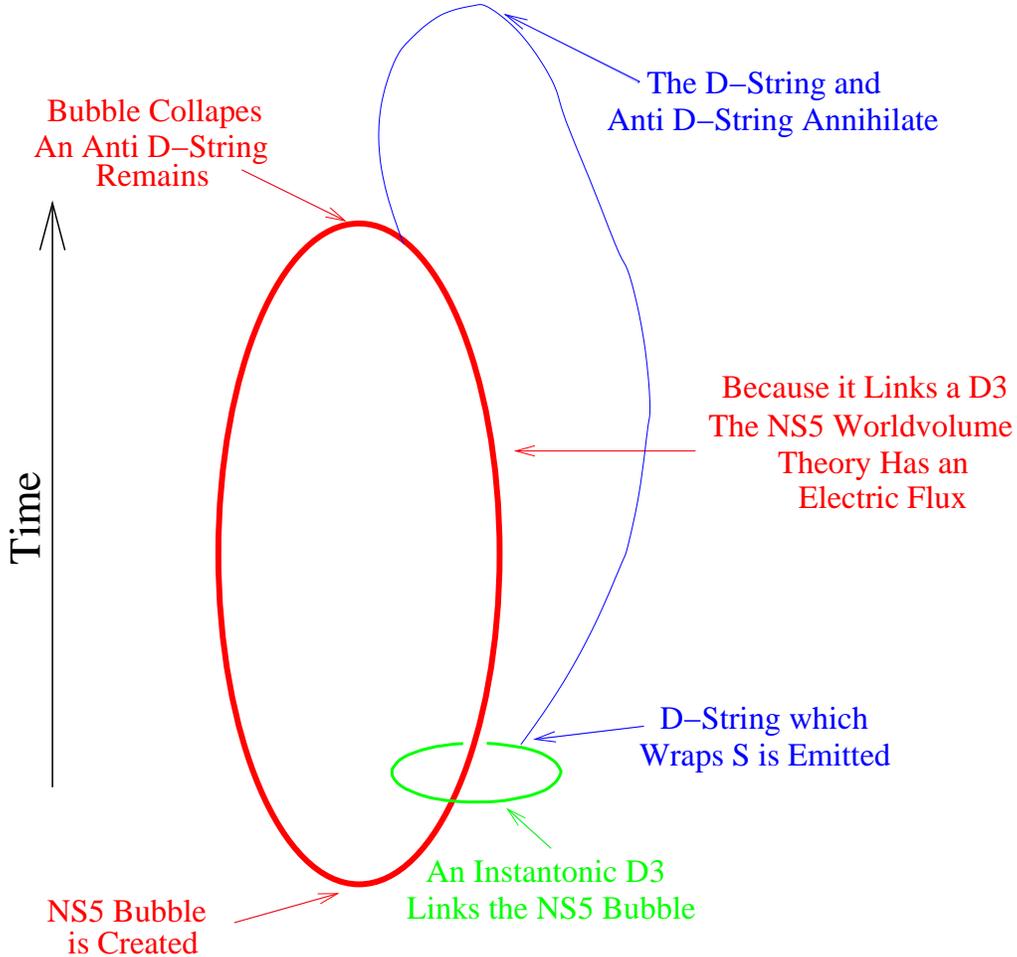}
\caption{The Second Process: An NS5-bubble spontaneously appears, wrapping the 1-cycle $S$.  Later an instantonic D3 appears linking the NS5 and also wrapping $S$.  The D3 decays to a D-string which wraps $S$.  The NS5 necessarily links the D3 and so it carries a topologically nontrivial dual fieldstrength $*F$.  To decay to nothing the NS5 bubble must first lose the charge corresponding to this dual fieldstrength, which corresponds to emitting an anti D-string.  In this picture the anti D-string later annihilates the D-string.  In the IR this reduces to Fig.~\ref{scen1}.}\label{scen2}
\end{figure}

In the second process, as depicted in Figure~\ref{scen2}, the NS5-brane bubble again forms from nothing and then grows.  An instantonic D3-brane may link this NS5-brane\footnote{An instantonic D3-brane which links the bubble may intuitively be thought of as being inside of bubble, for example in the case of IIB on $S^3$ the 3-spheres that link are on the inside of the 5-brane domain wall if the 3-sphere directions are projected out.} and also wrap a loop $S\in H_1(M^9)$.  When the instantonic 3-brane decays a D-string wrapping $S$ will be left over, as well as some number of other D-strings wrapping $S$ depending on the wrapping of the D3 over the 3-cycles in $M^9$ that do not link the NS5-brane.  For example if the D3 wraps $Z^3$ once then there will be $k$ additional D-strings.  However the additional D-strings carry trivial charge because they may be annihilated by D3 instantons that do not link an NS5-brane.  Thus the charge for D-strings wrapping $S$ has apparently increased by 1 unit, violating the desired charge conservation.  In the IR the D3-brane disappears into the NS5-brane and this D-string is emitted directly from the NS5, and thus this scenario reduces to the previous scenario.  

In fact, these two scenarios have the same resolution.  The D3 links the NS5 and so aquires a magnetic charge in its worldvolume gauge theory.  This magnetic charge is D-string charge in IIB supergravity and so when the D3 decays an extra D-string remains.  Similarly the NS5 links the D3 and so acquires a unit of electric charge in its worldvolume gauge theory, just as in the previous example.  Therefore just as in the previous example, this electric charge contributes to the D-string charge.  Again if the NS5 decays, an anti D-string wrapped on $S$ remains and the charge is therefore conserved.  As in the first case, this result is S-duality covariant and could also be applied to instantonic 5-branes to learn that they do not violate D3 charge conservation.

\subsection{Phase Transitions}

Thus it appears that the above interpretation of the 3-form fieldstrengths in the differentials (\ref{homseq}) allows the conserved charges to remain conserved, even in the presence of bubbles.  However inside of a bubble it appears as though the charge group has changed, it is only when the charges on the bubble are considered as well that the original charge group is conserved.  This means that if a bubble is allowed to extend to spatial infinity (or more generally to be non-compact) then the original charge may escape to infinity in finite time and so not be conserved.  A particularly simple example of this is a Brown-Teitelboim scenario \cite{BT,BT2}, which has been embedded in string theory by Bousso and Polchinski \cite{BP}. 

We will now investigate this possibility in an example.  For concreteness we will consider a spacetime of the form $\R\times S^3\times M^6$, where $\R$ refers to the time direction.  5-brane bubbles not extended along the $S^3$ may nucleate and in their interior the 3-form flux on the 3-sphere will be different from that on the exterior by 1 unit.  If the bubbles all decay then the charge at the end will be the same as the charge at the beginning.  However instead of decaying the bubbles may coalesce and at same point in time all of $M^9$ may be on the interior, changing the background flux on the 3-sphere.

This may be seen concretely by considering the above ``first process''.  An instantonic 5-brane may wrap $M^6$ and emit a string which wraps some 1-cycle in $M^6$.  This creates an electric flux on the 5-brane worldvolume.  If $M^6$ is compact then this 5-brane shrinks to a point before disappearing, with everything except for the point being the interior region where the flux has been changed.  This point carries the anticharges of all of the strings that have been emitted and so it decays into an antistring which exactly cancels the string that was emitted earlier.  The old conservation law is preserved in the sense that, were another 5-brane to be created and sweep out $M_6$ in the opposite direction, then the charge group would be the same as it was originally and the charge would be unchanged because the two instantonic 5-branes would link any instantonic 3-branes that had tried to violate charge conservation while the background flux was different.  Considering the IR, the same must work for the second process.  Intuitively the charge was conserved because the electric flux ``had nowhere to go'' in the compact 5-brane worldvolume, and so the total electric charge on the worldvolume must vanish, requiring an anti-string for every string created.

If $M^6$ is noncompact then it may host a net electric charge and so D-strings may be emitted freely and the prebubble conservation law is violated.  Once the region at infinity is in the bubble, that is once the boundary conditions have been changed by an infinite number of nucleating bubbles, then there is a new conservation law.  The cardinality of the charge group for D-strings has shifted by one.  Also the charge group for D3-branes will change as a result of the $H$-flux shifting by one unit.  Although the change in flux may occur in finite time, the escape of charge to infinity may require infinite time, in which case this would be the usual breakdown of charge conservation in a noncompact space.

To summarize, the 3-form fieldstrengths in the homology spectral sequence (\ref{homseq}) may be chosen arbitrarily, but they only yield the brane Page charge on a Cauchy surface when they are chosen to be the actual values of the fluxes on that surface.  If $M^9$ is compact then a bubble may eventually fill space and change the 3-form flux.  However if another bubble returns the flux to its original value then the total charge will agree with its original value.  In this sense charges are always conserved when time-slices are compact. 

If $M^9$ is noncompact and physical processes cause the boundary conditions to change, then the charge group will change as well.  In particular if there is a net 5-brane charge caused by the charge of a 5-brane domain wall that extends to infinity then the boundary conditions are not fixed and the prescription given here for classifying charges yields conserved charges but does not count branes.  On the other hand we have seen above that a compact 5-brane does not lead to such problems and that 1-brane number plus the 5-brane electric flux $\int *F$ is a conserved quantity.  In this case we are free to choose the fluxes around a 3-cycle with any 5-brane linking number so long as conventions are mutually consistent.

\section{D9-Branes and Speculations} \label{specsec}

\subsection{Dielectric D9-Branes}
We do not attempt to classify D7-branes because objects with D7-brane charge may have S-duals that lead to monodromies which generically violate our weak coupling assumption.  In addition the logarithmic falloff of energy as a function of distance from a D7-brane implies that the local geometry must be compact.  As the classification of branes in this note is treated independently from the classification of geometries, stronger tools may be required for such a persuit.  These two problems do not impede our study of dielectric D7-branes which carry no D7-brane charge, for example D7-branes that wrap a contractible 2-cycle $\Sigma$.  These have fieldstrengths with first Chern classes whose integral leads to D5-brane charge.  This is the reason that D5-brane charge in this paper is valued in finite cyclic groups.

In fact, we may even consider dielectric D9-branes.  10-dimenional gravitational anomaly cancellation proceeds as usual so long as such a dielectric D9-brane contains no D9-brane charge.  One example of a ``D9-brane'' with no D9 charge is a stack of spacefilling D9-$\overline{\textup{D9}}$ pairs.  In Ref.~\cite{Ktheory} Witten motivated the K-theory classification with the fact that the lower D-brane charges on such stack are classified by the K-theory of pairs of the gauge bundles on the D9's and $\overline{\textup{D9}}$'s respectively.  The K-theory equivalence relation comes from the D9-$\overline{\textup{D9}}$ annihilation studied by Sen in Ref.~\cite{Sen}.  In fact the instability of brane configurations in the image of $d_3$ was first described in Ref.~\cite{DMW} using this language.

Such a stack is just a special case of the MMS instantonic processes considered in this note.  D9-branes that carry no D9 charge do not need to be spacefilling.  On the contrary, a D7-brane on an 8-manifold $M^8$ may expand into a dielectric D9-brane which is actually a D9-$\overline{\textup{D9}}$ pair on $M^8\times B^2$ where $B^2$ is a 2-manifold transverse to the $M^8$.  T-duality seems to imply that low energy configurations will have a lot of gauge flux concentrated around the non-smooth edge.  To an observer such a brane describes the process of a D9-$\overline{\textup{D9}}$ pair being created at a point, growing in all directions, shrinking back to a point and then annihilating.  During the entire process the brane and anti-brane share a common boundary.  D9 charge vanishes everywhere and so again the gravitational anomaly is cancelled.  Sen's case is one in which the 2-manifold extends over all of the transverse space.  Notice that the same construction may be performed with any D$p$-brane and a $(9-p)$-manifold along which it expands to become a dielectric D9.  

It is possible that these non-spacefilling D9-branes, being deformations of lower D$p$-branes, are more amenable to S-duality than their spacefilling cousins.  In particular they may shed light on the nature of the desired generalization of K-theory via, for example, a
\begin{equation}
S_{D9}\supset *F\wedge B 
\end{equation}
term in the D9 action which allows fundamental string charge to be interpreted as a feature of the D9-brane's (dual) gauge bundle.  Could the generalization of K-theory be a classification on the category of bundles with a Hodge star?  If so a first place to look for such a K-theory might be in $N=4$ super Yang-Mills in 4 dimensions, where $*F$ has a geometric interpretation as the fieldstrength of another bundle.  Equivalently, it may be easier to understand what the generalized sequence computes if we first compactify on a $T^6$ so that this $*F$ term is a 2-form.  Note that such a compactification enlarges the S-duality group to $E_{7,7}(\Z)$ and so the corresponding generalization of K-theory may have a much larger covariance than $SL(2,\Z)$, such as $E_8$.  On the other hand, perhaps such a dimensional reduction would obscure a crucial part of the structure.

The existence of such dielectric branes implies that the $Dp$-brane in IIB has some mode for expanding in the transverse directions.  This mode has a potential, as in Ref.~\cite{BDS}, describing for example the energy required to have a finite radius.  This leads to two more conjectures.  The first is that for D-branes at sufficiently high temperature this mode is excited and branes are smeared in all transverse directions.  The second is that the quantum 0-point energy of the bosonic modes of expansion in the transverse direction give all D-branes an inherent thickness in the transverse directions.  This thickness may be smaller than the scale at which such notions are physically meaningful.

\subsection{The K-Theory of Dual Pairs}

In this subsection we turn off the background fields.  We demonstrate that in the absence of NS charges the classification of U(1) gauge fields on the dielectric D9-branes above suggests that lower-dimensional brane charges are classified by K-theory.  This construction extends naturally to include fundamental strings in a new structure which appears to be a K-theory of pairs of bundles that are Hodge dual in a sense to be made more precise momentarily.  Unfortunately the new structure may not yet be able to accomodate 5-branes, which may be a consequence of the fact that NS5-branes lead to a non-torsion $H$ field on the 3-sphere that links them and so cannot exist inside the worldvolume of D9-branes \cite{Kapustin}.  The inclusion of 5-branes may need to await a twisted version of this K-theory, which perhaps may be constructed along the lines of the geometric construction of Rosenberg's twisted K-theory in Ref.~\cite{Bgerbe}.  We will call the construction the K-theory of dual pairs, but we will neither prove nor suggest that it is a K-theory.

Consider a D-instanton at a smooth point in spacetime.  Although in the absence of background fields the lowest energy configuration of the instanton is pointlike, it may instead be dielectric.  For example it may expand into a D-string which wraps a contractible 2-sphere over which its gauge bundle has a first Chern class that integrates to one.  Although no dynamical process takes such a D-string to a D-instanton, a Gauss' law definition of charges does not distinguish them and neither can an observer who is only able to probe large distances.  In particular, the D-string may be continuously deformed to the D-instanton.  

We will be interested in two higher-dimensional versions of this phenomenon.  In the first, a dielectric D-instanton may wrap a contractible $S^2\times S^2$ in which case it is a D3-brane with only D-instanton charge and the D-instanton charge is determined by the second Chern character of its gauge bundle.  In the second, such a D3 may polarize yet further so as to wrap 
\begin{equation}
T=S^2\times S^2\times S^2\times S^2\times S^2
\end{equation}
in which case it is a D9-brane.  Such a D9 contains a worldvolume U(1) gauge theory and the fifth Chern character yields the D-instanton charge.  

Notice that, unlike the case of the lower dimensional dielectric branes, $T$ cannot be contractible because the deformation retract would sweep out an 11-dimensional space and we have only 10-dimensions at our disposal.  Thus if we insist on a contractible dielectric brane then it must have the topology of the 10-ball $B^{10}$ and the dielectric D9 must be projected into this 10-ball via a map
\begin{equation}
f:T\longrightarrow B^{10}.
\end{equation}
Notice that the map $f$ cannot be 1 to 1.  The preimage of each point in $B^{10}$ must contain at least 32 points in $T$ and precisely half of these must be mapped with each orientation.  For simplicity we will consider only the maps $f$ such that the preimage of each point in $B^{10}$ consists of a fixed number $n$ points.  The image of $f$ then describes the actual configuration of the D9 in spacetime.  The fact that the map is $2n$ to $1$ means that the D9 is folded into $n$ D9-branes and $n$ $\overline{\textup{D9}}$-branes, where $n\ge 16$.  The U(1) gauge bundle on the original D9 is then a U$(n)\times$U$(n)$ gauge bundle over $B^{10}$.  The lower-dimensional brane charges are still obtained by integrating the Chern characters of this gauge bundle.  

\begin{figure}[ht] 
  \centering \includegraphics[width=5.5in]{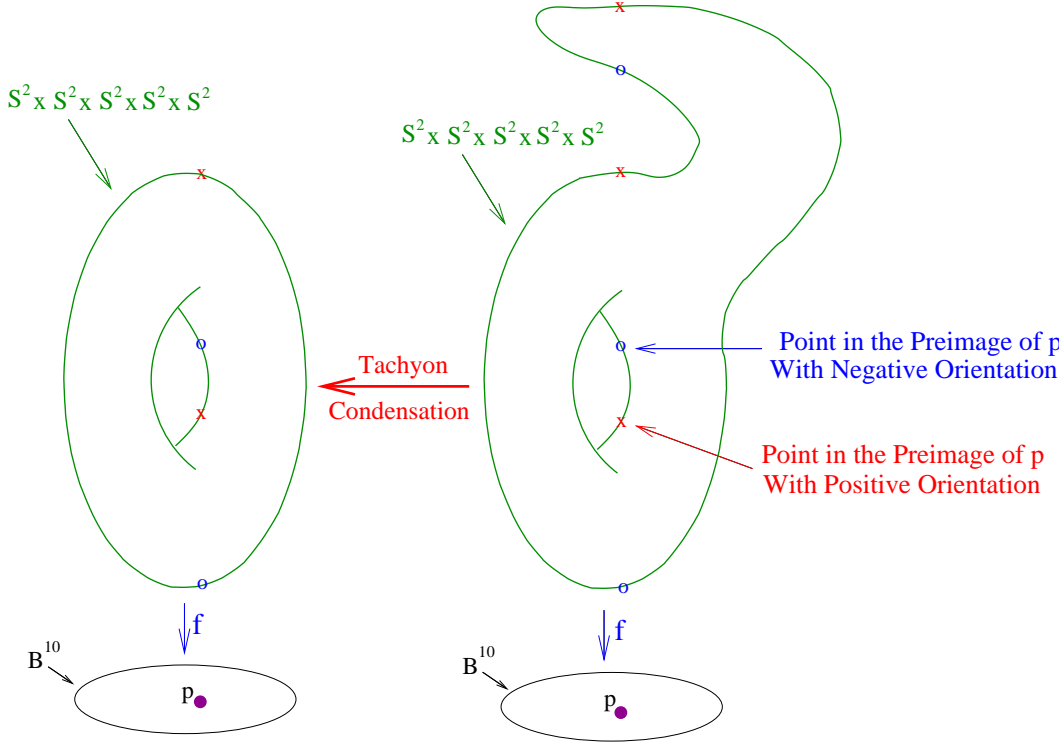} 
\caption{Above are two maps from $T=S^2\times S^2\times S^2\times S^2\times S^2$ to $B^{10}$.  Each point has a preimage which consists of $2n$ points, $n$ with each orientation.  The map on the right has a greater value of $n$ and decays via tachyon condensation to that on the left.  A choice of $U(1)$ bundle on $T$ yields a $U(n)\times U(n)$ bundle on $B^{10}$ for each $f$, all of which are in the same K-theory equivalence class.}\label{equivfig}
\end{figure}

The charges of the D9 are independent of the homotopies of $f$ which add an extra two sheets to the preimage, changing $n$ to, for example, $n+1$.  Such a deformation changes the gauge bundle above the 10-ball
\begin{equation}
U(n)\times U(n)\longrightarrow U(n+1)\times U(n+1).
\end{equation} 
In particular, if we diagonalize the bundles as we are free to do over a contractible space, such a deformation serves to direct sum an arbitrary U(1) bundle to each of the U($n$) components and these two U(1) bundes must be equivalent if the homotopy is to be continuous.  Thus a change in the map $f$ generates an equivalence relation on the bundles above the 10-ball
\begin{equation}
(E,E\p)\sim(E\oplus F,E\p\oplus F). 
\end{equation}
Two maps related by such a deformation may be seen in Figure~\ref{equivfig}.

The group of bundles on $B^{10}$ with this equivalence relation is refered to as the K-theory $K^0(B^{10})$.  If we wish to obtain the K-theory of a more complicated space, we may do so by dividing the space into such balls and then gluing them together.  Thus we recover Witten's \cite{Ktheory} classification of D-branes in IIB by $K^0$ of the spacetime.

Such a construction is readily generalized to include fundamental strings.  Fundamental strings may exist within a D9-brane worldvolume and are Poincare dual to $*F$.  In what follows we will make use of S-duality, and so for safety we will eliminate D5 and D7-branes from the spectrum by hand by restricting our bundles to have vanishing first and second Chern characters.  Such a restriction is quite unnatural as characteristic classes do not in general completely specify bundles, and so it is hoped that either a more natural restriction exists or that the K-theory of dual pairs is only natural when background fluxes are included and it is twisted, in which case 5-branes naturally appear as domain walls in these fluxes.

Once again we begin the D3-brane.  If this D3-brane has an image which contains a 2-cycle then it may contain both F-string and D-string fluxes.  The D-string is Poincare dual to the U(1) gauge fieldstrength $F$ while the F-string is dual to the U(1) fieldstrength of the dual gauge bundle, $*F$.  The image of maps of such a D3-brane onto a the 4-ball with gluing rules for the 4-balls to allow string charge may generate a 4-dimensional version of the K-theory of dual pairs.  

However we will be interested in the 10-dimensional version.  To construct this we repeat the construction above and expand the D3 into a dielectric D9 wrapping $T$.  By construction we allow D-instanton, string and D3 charges, although the string and D3 charges will only be present after patches are glued.  In particular the charges are given by Chern characters
\begin{equation}
Q_{D(-1)}\sim\int F^5\sp
Q_{D1}\sim\int F^4\sp
Q_{F1}\sim\int *F\sp
Q_{D3}\sim\int F^3.
\end{equation} 
Alternately we could have S-dualized the dielectric D3 before blowing it up into a dielectric D9.  This would have again yielded a dielectric D9, but the charges would be different and so it would have a different gauge bundle with a different fieldstrength $G$, which would be related to the original charges by
\begin{equation}
Q_{D1}\sim\int F^4\sim\int *G \sp
Q_{F1}\sim\int *F\sim\int G^4\sp
Q_{D3}\sim\int F^3=\int G^3.
\end{equation} 
In particular, taking the Hodge dual of the equivalence for D-strings (which became F-strings after the S-duality) one learns that up to a gauge transformation
\begin{equation} \label{pair}
G\sim*F^4.
\end{equation}
Thus the generalization of K-theory that seems to include fundamental strings is not a K-theory of bundles, but rather a K-theory of pairs of dual bundles related by Eq.~(\ref{pair}) and whose third Chern characters agree.  Again it would be preferable if dual bundles could be defined without reference to characteristic classes to avoid potential ambiguities when the characteristic classes miss crucial topological information.  In a perfect world the 5-branes appear as the second Chern characters of the two bundles.

Again to get actual D9-branes in spacetime one must map $T$ to $B^{10}$.  Homotopies of maps will again yield equivalence relations on the two bundles.  To get nonvanishing charges one must then glue $B^{10}$'s together with gluing rules.  This K-theory will turn out to be finer than the usual K-theory because the homotopies must respect the fact that both $F$ and $G$ are fieldstrengths of bundles, and so less configurations will be identified.  The fact that this K-theory is finer reflects the fact that it distinguishes states with different numbers of F-strings, whereas the usual K-theory does not.

Extending the usual explicit constructions of such bundles may be possible in this case as the 32-sheeted preimages lend themselves well to $\Gamma$-matrices $\Gamma_i^\pm$ whose eigenvalues on the $i$th sphere are plus and minus one depending on the orientation with which it is mapped.

\section{Comparison with IIB on the 3-Sphere} \label{compsec}

\subsection{Spectra as Computed Above}
The spectrum of D-branes on the product of a contractible 7-manifold and the 3-sphere was calculated with no knowledge of the nature of the contractible space.  As a result the kinds of charges that may exist there were undetermined.  For example it could be an AdS space, in which case objects extended in the $AdS$ directions have conserved charges because they end on D-branes on the horizon.  Charges could also be acquired by terminating on an end of the world like those of Ref.~\cite{HW}.  In general each type of allowed boundary condition provides some type of charge.  If we allow the 7-manifold to be noncontractible, although to factor into $\R_{time}\times M^6$ with $M^6$ orientable, little would have changed but the allowed charges would have been extended to include wrappings of the cycles on $M^6$.  The actual charge group is then the tensor product over $\Z$ of the charge group for $M^6$ and the charge group computed above for the 3-sphere.  That is to say, a basis of branes consists of objects which have a nonvanishing $M^6$ charge and also a non-vanishing charge taken from Table 1.

To compare the spectrum with the result of the generalized spectral sequence we will need to choose a set of charges for objects on $M^6$.  We will denote the group of charges for an object extended in $p$ directions along $M^6$ by $J_p$.  This is the group of charges that come from, for example, the nontrivial homology of $M^6$ or from ending on a D3-brane on the horizon of $AdS^5$.  With this notation we may now summarize the charges computed above:

\vspace{.3in}
\hspace{1in}
\begin{tabular}{c|c}
Charge&Charge\\
Type&Group\\ \hline
F1&$Z_j\otimes_\Z J_1$\\
D1&$Z_k\otimes_Z J_1$\\
D3&$Z_{gcd(j,k)}\otimes_Z J_3$\\
D5&$Z_k\otimes_Z J_5$\\
Vertex&$\Z\otimes_Z J_0$\\
\multicolumn{2}{c}{}\\
\multicolumn{2}{c}{Table 2: Charge Groups on $S^3$}
\end{tabular}
\vspace{.3in}

More general boundary conditions, which assign different groups to NS and RR charged objects, may exist in some examples (such as IIB string theory on $AdS^3\times S^3\times T^4$).  In such examples S-duality invariance is explicitly violated by boundary conditions and so the S-duality covariant AHSS would also need to be modified, at least by modifying the generalized homology defined in Eq.~(\ref{hom}). 

\subsection{Spectra as Computed by the Generalized AHSS}

Now we want to reproduce these results using the generalized spectral sequence developed above.  The first step is to compute the generalized homology groups.  We consider manifolds of the form $\R\times M^9$.  In the simplest case $M^9$ is compact and the groups of interest are the wrappings of solitons on the 9-manifold.  In this case $M^6$ would also be compact and therefore not contractible.  The charges of branes wrapped on $M^6$ are just the homologically inequivalent cycles that they may wrap and so the homology of the 6-manifold would be $H_p(M^6)=J_p$.    

In the non-compact case we need to define what kind of homology we wish to use.  To get the right answer we will define $H_p(M^6)=J_p$.  While in the compact case this is the usual integral homology, in the noncompact case it is something very strange that knows about, for example, different types of branes that a charge may end on at infinity.  However such an unusual definition is necessary as the classification of charges (some of which will generally extend to spatial infinity) on a noncompact space does depend on such boundary conditions.   We will restrict our attention to orientable 6-manifolds, and so 
\begin{equation}
J_6=H_6(M^6)=\Z.
\end{equation}
We define the homology groups of the time direction to be
\begin{equation}
H_0(\R)=H_1(\R)=\Z .
\end{equation}
It would, however, be possible to proceed with this calculation using only the $M^9$ slice of spacetime.  Combining these expressions we find
\begin{equation}
H_7(\R \times J_6)=H_1(\R)\otimes_\Z H_6(M^6)=\Z.
\end{equation}
We will refer to the generator of this group as $\alpha$.

The homology of each timeslice $M^9$ of spacetime is then
\begin{equation} \label{s3hom}
H_p(M^6\times S^3)=\sum_{i}H_i(M^6) \otimes_\Z H_{p-i}(S^3)=J_p\oplus J_{p-3} 
\end{equation}
Intuitively cycles that do not wrap the sphere are classified by $J_p$ and those that do by $J_{p-3}$.  The Poincare duals of the background fieldstrengths are
\begin{equation}
PD(G_3)= j\alpha\sp
PD(H)= k\alpha.
\end{equation}
Now we may finally calculate the maps $d_3$.

Recall that $j$ and $k$ are taken to be nonvanishing.  By antisymmetry of the cap product, the cap product of $\alpha$ with a cycle $x$ is nontrivial only if $x$ is a product of another cycle and the generator of $H_3(S^3)=\Z$.  Therefore the kernel of $d_3$ acting on a $p$-cycle only receives contributions from the $J_p$ term on the RHS of (\ref{s3hom}).  Physically this means that the instantonic branes are precisely those that wrap the $S^3$, corresponding to the $J_{p-3}$ term.  

The group of charges consists of equivalence classes of the elements of $\ho\in$\ ker$(d_3)$.  Let $ker_p$ be the $p$-dimensional part of the kernel of $d_3$.  Then applying (\ref{homseq}) to $\ho$ we have computed
\begin{equation}
ker_1=(J_1,J_1)\sp
ker_3=J_3\sp
ker_5=(J_5,J_5).  
\end{equation}
We need to quotient this by the image of the map $d_3\p:\he\longrightarrow\ho$.  Again this map acts nontrivially precisely on the $p$-cycles that are products of the 3-cycle in $S^3$ and another $(p-3)$-cycle, the $J_{p-3}$ terms.  Due to the cap product with $\alpha$, the image of these cycles is not a product with the generator of $H^3(S^3)$ and so is a $J_p$ term.  These $J_p$ terms are multiplied by $j$ or $k$ because the Poincare duals of $G_3$ and $H$ are $j\alpha$ and $k\alpha$ and the cap product is bilinear.  If we denote the $p$-dimensional part of the image by $im_p$ then we find
\begin{equation}
im_1=(k J_1,j J_1)\sp
im_3=kJ_3+jJ_3=gcd(j,k)J_3\sp
im_5=kJ_5.
\end{equation}
There is no second component in $im_5$ because we do not know the part of $d_3$ that takes 7-brane charges to $NS5$-brane charges.  This entry is simply omitted in the definition of the sequence (\ref{homseq}).  

Assembling the known kernels and images we can compute the charge group $Q_p$ for objects extended in $p$ spatial dimensions
\begin{equation}
Q_1=(\Z_k\otimes_\Z J_1,\Z_j\otimes_\Z J_1)\sp
Q_3=\Z_{gcd(j,k)}\otimes_\Z J_3\sp
Q_5=(\textup{Unknown},\Z_k\otimes_\Z J_5).
\end{equation}
This is in perfect agreement with Table 2, except that the modified spectral sequence missed the vertices.  The vertices are missing because they were explicitly thrown away.  The vertex states are the elements of $\ho$ that are not in the kernel of $d_3$, and so must be the endpoints of other solitons.  No vertex instability is found in the first differential, $d_3$, but vertex decay may be expected whenever the composition of two differentials is nontrivial.

\section{A Conjecture for Non-$spin^c$ Manifolds} \label{nonspincsec}

Finally we turn our attention to the classification of branes on non-$spin^c$ manifolds.  Due to dimensional constraints in IIB this generalization only affects the action of $d_3$ on the 5-brane charges, no other branes may wrap non-$spin^c$ cycles.  The most naive modification comes from applying the exterior derivative to the corresponding cohomology formula and Poincare dualizing:
\begin{equation} \label{guess}
d_3(D_5,NS_5)=PD(i^{D5}_*(H+W_3(D5))+i^{NS5}_*(G_3+W_3(NS5))).
\end{equation}
Here $W_3(D5)$ and $W_3(NS5)$ are the third Stiefel-Whitney classes of the worldvolumes of the disjoint union of all D5 and NS5-branes respectively.  The maps $i^{D5}_*$ and $i^{NS5}_*$ increase the dimension of a cohomology class by 4, they are pushforwards by the inclusion maps of the D5 and NS5-brane worldvolumes into spacetime.

Fortunately the soliton spectrum of a non-$spin^c$ example, string theory on $AdS^5\times\rp^5$, has already been worked out in full \cite{baryonz}.  The sequence needs slight modification for a test against this example because, as an orientifold, some charges inhabit twisted homology groups instead of ordinary homology groups.  The background $G$ and $H$ are $\Z_2$-valued discrete torsions and the decay of an instantonic 5-brane into 3-branes is described instead in terms of vertices.  We recall from the $S^3$ examples that for each kind of vertex there is an instantonic-brane decay processes that may be obtained by rotating the vertex so that it does not continue into the infinite past or future while the anomaly-cancelling branes are rotated so that they do continue into the infinite past or future.  While there may be topological obstructions to such a process in general, there is no obstruction in the present example.  

In the presence of discrete fluxes a vertex must be the endpoint of a lone additional 3-brane, which in the instantonic brane picture is precisely the 5-brane to 3-brane decay that the relevant component of $d_3$ computes.  More precisely, it is found that an instantonic D5 decays to an even number of $D3$'s when there is 1 unit of $H$ flux and an odd number when there is no $H$ flux.  By S-duality the same is true of an $NS5$ with $G_3$ flux.  

This may be compared with the guess Eq.~(\ref{guess}).  The RHS is valued in $\Z_2$.  For a $(p,q)$ 5-brane with 3-form flux $(j,k)$ the guess yields
\begin{equation}
d_3(q,p)=qk+q^2+pj+p^2.
\end{equation}
This is manifestly $SL(2,\Z)$ invariant.  If $(p,q)=(0,1)$ as in \cite{baryonz} then the RHS is $k+1$.  We recall that this is to be interpretted modulo $2$ and gives the number of remaining D3's.  Thus there will be a remaining D3 precisely when the $H$ discrete torsion is turned off, ie for SO($N$) gauge theories.  This is the correct answer, but is only one example.  An analysis of the SU(3) group manifold, which has a non-$spin^c$ submanifold on which branes wrap, should provide another test. 

\section{Conclusions}
We have proposed an extension of the AHSS which we hope will compute the charges of strings and branes in IIB on the spacetime $\R\times M^9$ and have tested it in the cases $M^9=M^6\times S^3$.  In fact we have produced only the first ``differential''.   The first thing that needs to be done with any such proposal is that it must be checked in more examples.  There are many simple examples where such a check may occur.  For example, the full soliton spectrum of IIB on $AdS^5\times S^5$ has already been evaluated in Ref.~\cite{baryonz}.  D-brane charges on the group manifold SU(3) were evaluated in Ref.~\cite{MMS} and the extension to other charges does not seem difficult.  

An obvious place to look which seems to be unexpectedly difficult is $\rp^3$, the group manifold of SO(3).  The WZW model on SO(3) contains a brane which wraps a 2-cycle.  However
\begin{equation}
H_2(\rp^3,\Z)=0\sp
H_2(\rp^3,\tilde{\Z})=\Z_2
\end{equation}
where the second homology class has coefficients in the twisted sheaf of integers.  This means that this D-brane lives in the twisted cohomology group and so this WZW model is realized by a geometry with an orientifold O5-plane.  The O5-plane is not S-duality invariant \cite{baryonz} and so the above strategy must be modified to consider this case, which is in many ways more the domain of type I string theory than IIB.

If these checks are successful, then one may try to fill in the major gaps in this proposal.  There should be at least one more differential, $d_5$, which describes for example $G_5$ flux on an instantonic D5-brane which causes it to decay to a collection of F-strings.  This extra differential will be required to get the correct spectrum for string theory on $AdS^5\times S^5$.  In addition there may be a $d_7$ which discribes the same effect with $*G_3$ flux on D7-branes.  However, as with the O5-plane, S-duality may often be broken in the presence of a D7-brane.

Once the sequence is complete the problem of how to formulate the relevant extension problem will remain.  However in examples this is often clear, and so perhaps some insight may be gained by studying solitons on yet more spaces.

\noindent 
{\bf Acknowledgements}

\noindent
I'd like to thank Allan Adams for comments and also the good people of the INFN for providing me with food, shelter, money and oil while this work was in progress.

\noindent

\bibliographystyle{ieeetr} 
\bibliography{ex}
\end{document}